\documentclass[twocolumn,aps,showpacs,superscriptaddress,tightenlines,a4paper]{revtex4}

\usepackage{amsmath}
\usepackage{paralist}
\usepackage{graphicx}
\usepackage{xcolor}

\begin{document}
%
%
	\title{Transformation optics, isotropic chiral media, and non--Riemannian geometry}
%
%
	\author{S. A. R. Horsley}
	\affiliation{School of Physics and Astronomy, University of St Andrews, North Haugh, St Andrews, KY16 9SS, UK}
	\email{sarh@st-andrews.ac.uk}
%
%
	\begin{abstract}
		The geometrical interpretation of electromagnetism in transparent media (transformation optics) is extended to include media with isotropic, inhomogeneous, chirality.  It is found that such media may be described through introducing the non--Riemannian geometrical property of torsion into the Maxwell equations, and shown how such an interpretation may be applied to the design of optical devices.
	\end{abstract}
%
%
%
%
%
%
%
%
        \pacs{81.05.Xj,02.40.Hw,78.20.Ek,42.25.Ja,42.15.-i}
	\maketitle
	\bibliographystyle{unsrt}
%
%
	\section{Introduction\label{intro}}
	Several authors have noticed that the free space Maxwell equations in an arbitrary co--ordinate system, on a Riemannian space--time background, take the same form as they do in a certain class of transparent, inhomogeneous, anisotropic media (see references in~\cite{leonhardt2006b} and also~\cite[\textsection{90}]{volume2}).  Recently this relationship has been put to work in reverse.  \emph{Transformation optics}~\cite{leonhardt2010,chen2010,leonhardt2006,pendry2006,shalaev2010} has developed to use this formal analogy to design optical devices, and even investigate analogues of astrophysical objects in the laboratory~\cite{philbin2008,belgiorno2010}.
	\par	
	Light propagation on a Riemannian space--time background can be intuitively understood in terms of rays following geodesics, and polarization undergoing parallel transport along each ray~\cite{leonhardt2010,born1999}.  Transformation optics uses this simple picture to design a space that acts on the optical field in a desired way, and, using the peculiar property of Maxwell's equations mentioned in the preceding paragraph, thereby determines the necessary material properties from geometrical quantities.  Notably, this recipe has been applied to derive the material properties necessary for devices that conceal objects from the electromagnetic field~\cite{pendry2006,leonhardt2009}, and focus light intensity into a region that is smaller than the diffraction limit leads us to expect~\cite{pendry2000,leonhardt2009b}.  In this paper, the idea is to provide more variables for this design strategy to explore: in particular, to incorporate chiral media into transformation optics.
	\par
	As stated above, the existing theory of transformation optics works within Riemannian geometry, where the `design parameter' is the space time metric, \(g_{\mu\nu}\).  Yet, the term transformation optics came from the initial use of co--ordinate transformations to arrive at material parameters---i.e. Euclidean geometry.  Therefore, in the original sense, transformation optics works through the specification of three functions of position.  Clearly the full Riemannian geometry has greater freedom, with a symmetric space--time metric containing ten independent functions of position, translating into nine independent material parameters~\footnote{The conformal invariance of Maxwell's equations (i.e. invariance under \(g_{\mu\nu}\to f(x^\sigma)g_{\mu\nu}\)) has the consequence that out of \(n\) independent space--time metric components there are \(n-1\) independent material parameters.}.
	\par
	However, even if we recognise that a geometry must affect electric and magnetic fields in the same way, then this may not be the full story.  Symmetric, impedance matched permittivity and permeability tensors (i.e. \(\boldsymbol{\epsilon}/\epsilon_{0}=\boldsymbol{\mu}/\mu_{0}\)) represent six independent quantities, and the possibility of magneto-electric coupling, at least another six components.  This counting argument leaves three real magneto-electric coupling parameters that cannot be represented within Riemannian geometry, but may well have a geometrical interpretation~\footnote{When considering complex entries in the \(\epsilon\), \(\mu\), and electric--magnetic coupling tensors, the number of under-described parameters is larger.}.  It is therefore worth investigating non--Riemannian extensions to transformation optics.  In such a geometry there is at least one additional field---the space--time \emph{torsion}, \({T^{\mu}}_{\nu\sigma}={\Gamma^{\mu}}_{\nu\sigma}-{\Gamma^{\mu}}_{\sigma\nu}\)---that may be freely specified independent of the metric.  Torsion has already been explored in other analogue systems: e.g. the theory of sound waves propagating through superfluids, where non--Riemannian geometry has been used to describe the interaction with vorticity~\cite{andrade2004,andrade2005}.
	\par	
	In the following it will be shown that if we couple a non--Riemannian geometrical background to the free space Maxwell equations in a certain way, then Maxwell's equations can be interpreted as if in an inhomogeneous, isotropic, chiral medium described both by a Tellegen parameter, \(\chi\) and a chirality parameter, \(\kappa\).  In the limit of geometrical optics, this coupling is shown to reproduce the usual geodesic and parallel transport equations, but in the presence of geometrical torsion: i.e. optical activity is shown to have a geometrical interpretation in terms of the torsion tensor.
%
%
	\section{The relationship between EM fields in continuous media and Riemannian geometry\label{riemann}}	
	We begin by reviewing the existing theory of transformation optics~\cite{ward1996,leonhardt2010,shalaev2010}.
	\par
	Consider a space--time that is not necessarily flat---i.e. where the curvature tensor, \({R^{\mu}}_{\nu\sigma\tau}\), may not vanish---and where the co--ordinates, \(x^{\mu}\), are arbitrary.  In this instance, we write down the free space Maxwell's equations using their usual four dimensional form~\cite{volume2}, along with the convention from general relativity~\cite{hurley2000}, that ordinary partial derivatives be replaced by covariant ones, \(\partial_{\mu}\to\nabla_{\mu}\), and that the permutation symbol be scaled by the volume element, \(e^{\mu\nu\sigma\tau}\to \epsilon^{\mu\nu\sigma\tau}=g^{-1/2}e^{\mu\nu\sigma\tau}\)~\cite{lovelock1975},
        \begin{align}
           \epsilon^{\mu\nu\sigma\tau}\nabla_{\nu}F_{\sigma\tau}&=0\label{maxwell-free-3}\\
           \nabla_{\mu}F^{\mu\nu}&=0.\label{maxwell-free-4}
        \end{align}
        The covariant derivative, \(\nabla_{\mu}\), in (\ref{maxwell-free-3}--\ref{maxwell-free-4}) differs from an ordinary partial derivative by a quantity, \({\Gamma^{\sigma}}_{\nu\mu}\), known as the connection symbol~\cite{lovelock1975}.  Transformation optics works because, in the Riemannian case, the connection symbol in (\ref{maxwell-free-3}--\ref{maxwell-free-4}) plays the same algebraic role as the difference between the field equations in vacuum and in a polarizable medium.  We can see this through explicitly identifying \({\Gamma^{\sigma}}_{\nu\mu}\) in (\ref{maxwell-free-3}) and (\ref{maxwell-free-4}),  
        \begin{align}
           \epsilon^{\mu\nu\sigma\tau}\left[\partial_{\nu}F_{\sigma\tau}-{\Gamma^{\alpha}}_{\sigma\nu}F_{\alpha\tau}-{\Gamma^{\alpha}}_{\tau\nu}F_{\sigma\alpha}\right]&=0\label{maxwell-free-5}\\
          \partial_{\mu}F^{\mu\nu}+{\Gamma^{\mu}}_{\alpha\mu}F^{\alpha\nu}+{\Gamma^{\nu}}_{\alpha\mu}F^{\mu\alpha}&=0.\label{maxwell-free-6}
        \end{align}
        The assumptions of Riemannian geometry lead to a particular form for \({\Gamma^{\mu}}_{\nu\sigma}\) that is known as a Christoffel symbol~\cite{lovelock1975}, \({\Gamma^{\mu}}_{\nu\sigma}=\left\{{}^{\;\mu}_{\nu\sigma}\right\}\), and depends only on the form of the metric tensor, \(g_{\mu\nu}\),
        \begin{equation}
            \left\{{}^{\;\mu}_{\nu\sigma}\right\}=\frac{1}{2}g^{\mu\tau}\left[\partial_{\sigma} g_{\nu\tau}+\partial_{\nu}g_{\tau\sigma}-\partial_{\tau}g_{\sigma\nu}\right].\label{christoffel}
        \end{equation}
        Due to the symmetry of the Christoffel symbol in its lower two indices, (\ref{maxwell-free-5}) is indifferent to the distinction between \(\nabla_{\mu}\) and \(\partial_{\mu}\).  \emph{Therefore the definition of the field tensor in terms of the vector potential is the same as if the co--ordinates were those of a Galilean system}~\footnote{Throughout we use the term Galilean in accordance with Landau and Lifshitz, i.e. to mean the system of co--ordinates where the metric equals, \(\eta_{\mu\nu}=\text{diag}(1,-1,-1,-1)\), everywhere.}.  Furthermore, the antisymmetry of the field tensor means that the final term to the left of the equals sign in (\ref{maxwell-free-6}) is also zero.  We can now see that the only term that distinguishes the Maxwell equations (\ref{maxwell-free-5}) and (\ref{maxwell-free-6}) from a Galilean system is proportional to the trace of the Christoffel symbol~\cite{lovelock1975},
        \[
            \left\{{}^{\;\mu}_{\alpha\mu}\right\}=\frac{1}{2}g^{\mu\tau}\partial_{\alpha} g_{\mu\tau}=\frac{1}{2g}\partial_{\alpha}g,
        \]
        where \(g=\text{det}(g_{\mu\nu})\).  The free space Maxwell equations on a Riemannian background, (\ref{maxwell-free-3}) and (\ref{maxwell-free-4}), can thus also be written in a very similar form to the free space Maxwell equations in Galilean space--time.  All that is changed is the appearance of a factor of \(\sqrt{-g}\), and the relationship between \(F^{\mu\nu}\) and \(F_{\mu\nu}\)~\footnote{The minus sign is introduced into the square root of \(g\) for the sake of convention: physical space--time has the signature \((1,3)\), and therefore a negative value for the determinant, \(g\).  However, transformation optics is free to explore media that are equivalent to space--times with arbitrary signature (e.g. see~\cite{smolyaninov2010}).},
        \begin{align}
            e^{\mu\nu\sigma\tau}\partial_{\nu}F_{\sigma\tau}&=0\label{final-maxwell-1}\\
          \partial_{\mu}\left(\sqrt{-g}F^{\mu\nu}\right)&=0.\label{final-maxwell-2}
        \end{align}		
        \par
        There are two equivalent ways to understand (\ref{final-maxwell-1}) and (\ref{final-maxwell-2})---either as we have done so far, in terms of an empty, possibly non--flat, space--time background; or equivalently in terms of a Galilean system, containing a dielectric medium.  For, if we consider the four dimensional Maxwell equations in the presence of a material medium, within a Galilean co--ordinate system, and without any external sources, then Maxwell's equations can be written as~\cite{volume8}, 
        \begin{align}
            e^{\mu\nu\sigma\tau}\partial_{\nu}F_{\sigma\tau}&=0\label{maxwell-material-1}\\
            \partial_{\mu}G^{\mu\nu}&=0\label{maxwell-material-2}.
        \end{align}
        In this case there are two separate, but not independent, `field tensors'; the tensor, \(F_{\sigma\tau}\)---a bivector containing the physical fields, \(F_{\sigma\tau}=(\mathbf{E}/c,\mathbf{B})\); and the tensor, \(G^{\mu\nu}\)---a bivector containing the material fields, \(G^{\mu\nu}=(-c\mathbf{D},\mathbf{H})\), the vanishing divergence of which indicates that the material has no net charge, and that no net current passes through any cross section.  The constitutive relationship between \(F_{\sigma\tau}\) and \(G^{\mu\tau}\) serves to indicate how the physical fields are influenced by the medium.
        \par
        As the form of (\ref{final-maxwell-1}--\ref{final-maxwell-2}), is identical to that of (\ref{maxwell-material-1}--\ref{maxwell-material-2}), there is a correspondence between each and every co--ordinate system on any Riemannian background and an equivalent material, described within a Galilean co--ordinate system,
        \begin{equation}
            G^{\mu\nu}\leftrightarrow\frac{\sqrt{-g}}{\mu_{0}}F^{\mu\nu},\label{material-tensor}
        \end{equation}
        where the factor of \(\mu_{0}\) has been introduced so that the units agree with the physical interpretation of \(G^{\mu\nu}\).  The co--ordinates, \(x^{\mu}\), on a general background, are re--interpreted as a Galilean system, \(g_{\mu\nu}\leftrightarrow\eta_{\mu\nu}\), with the contravariant field tensor appearing as a material field.  The constitutive relationship between the physical fields, \(F_{\mu\nu}\), and the material fields, \(G^{\mu\nu}\) is obtained from the relationship between covariant and contravariant indices on the Riemannian background, 
        \begin{equation}
            F_{\mu\nu}=g_{\mu\sigma}g_{\nu\tau}F^{\sigma\tau}=g_{\mu\sigma}g_{\nu\tau}\frac{\mu_{0}}{\sqrt{-g}}G^{\sigma\tau}\label{constitutive},
        \end{equation}
        To make the interpretation more transparent, we introduce the usual three dimensional quantities; \(E_{i}=F_{0i}\); \(B_{i}=-\frac{1}{2}e^{ijk}F_{jk}\); \(cD_{i}=G^{i0}\); and \(H_{i}=-\frac{1}{2}e_{ijk}G^{jk}\), and adopt dyadic notation, so that (\ref{constitutive}) becomes,
        \begin{align}
            \mathbf{D}&=\boldsymbol{\epsilon}\cdot\mathbf{E}+\frac{1}{c^{2}}\mathbf{g}\times\mathbf{H}\label{consD}\\
            \mathbf{B}&=\boldsymbol{\mu}\cdot\mathbf{H}-\frac{1}{c^{2}}\mathbf{g}\times\mathbf{E}\label{consB2}
        \end{align}
        where, \(\epsilon_{ij}=\frac{\epsilon_{0}\sqrt{-g}}{g_{00}}\gamma_{ij}^{-1}\), \(\mu_{ij}=\frac{\mu_{0}\sqrt{-g}}{g_{00}}\gamma_{ij}^{-1}\), and we have introduced the symbols; \(\gamma_{ij}=(g_{0i}g_{0j}/g_{00} - g_{ij})\); \(\gamma^{-1}_{ij}=-g^{ij}\); and \(g_{j}=c g_{0j}/g_{00}\).  The conclusion is therefore that each co--ordinate system, on every Riemannian background, can be considered to appear to the electromagnetic field as a continuous medium with equal relative permeability and permittivity tensors (an impedance matched medium)---\(\epsilon_{ij}/\epsilon_{0}=\mu_{ij}/\mu_{0}\)---described within a Galilean system of co--ordinates.
	\par
	Perhaps even more importantly, the reverse also holds; for a fixed frequency of the electromagnetic field, all transparent, impedance matched media can be understood in terms of the vacuum Maxwell equations within a Riemannian geometry.  However, this space--time co-ordinate system is not uniquely defined by the medium, due to the invariance of the Maxwell equations under conformal transformations. 
%
%
        \section{Isotropic chiral media\label{chiral}}
        \par
        Section \ref{non-riemann} will describe a class of non--Riemannian geometries that are equivalent to inhomogeneous, isotropic, chiral media.  Unfortunately, there seems to be no agreement on the form of the constitutive relations that should be used to describe such media.  Although, in the frequency domain, these various constitutive relations can be shown to be \emph{physically} equivalent~\cite{lakhtakia1994,lindell1994}, the meaning of the individual material parameters \emph{is} different for each constitutive relation.  This will turn out to be important when we come to interpret chiral parameters in terms of geometrical quantities.  Therefore we spend this section distinguishing the constitutive relations before, in the next, introducing the geometry.
	\par	
	Landau and Lifshitz determine the relationship between the material fields and the physical fields in optically active media in terms of anti--symmetric complex components in the permittivity tensor: \(\epsilon_{ij}=\epsilon_{ji}^{\star}\)~\cite{volume8}.  This is equivalent to what is known as the Drude--Born constitutive relation~\cite{lakhtakia1994,lindell1994}, where \(\mathbf{D}\) is coupled not only to \(\mathbf{E}\), but also to \(\boldsymbol{\nabla}\times\mathbf{E}\).  For a geometric interpretation, we must have constitutive relationships that are local and have symmetric coupling terms in \(\mathbf{D}\) and \(\mathbf{B}\) (or \(\mathbf{H}\))~\footnote{The Drude--Born--Federov constitutive relations modify the Drude--Born relations to describe media with symmetric magneto-electric coupling (see e.g.~\cite[\textsection{3-2}]{lakhtakia1994}), coupling \(\mathbf{D}\) to \(\boldsymbol{\nabla}\times\mathbf{E}\) as well as \(\mathbf{B}\) to \(\boldsymbol{\nabla}\times\mathbf{H}\).  Clearly, these relationships are also non--local.}, and so exclude this possibility.  Instead, we describe isotropic chirality via the electric--magnetic coupling terms that were discussed in the introduction.  The Tellegen constitutive relations are often used~\cite{lakhtakia1994},
        \begin{align}
            \mathbf{D}&=\boldsymbol{\epsilon}_{\text{\tiny{T}}}\cdot\mathbf{E}+\frac{1}{c}\left(\chi_{\text{\tiny{T}}}-i\kappa_{\text{\tiny{T}}}\right)\mathbf{H}\label{chiral1}\\
            \mathbf{B}&=\boldsymbol{\mu}_{\text{\tiny{T}}}\cdot\mathbf{H}+\frac{1}{c}\left(\chi_{\text{\tiny{T}}}+i\kappa_{\text{\tiny{T}}}\right)\mathbf{E}.\label{chiral2}
        \end{align}
        where \(\chi\) represents the Tellegen parameter~\cite{tellegen1948}, and \(\kappa\) the chiral parameter.  The presence of the \(i\) indicates that, so long as \(\kappa\neq0\), (\ref{chiral1}) and (\ref{chiral2}) only have a meaning in the frequency domain.  While the existence of media with non--zero \(\kappa\) is unquestionable, there is a history of debate regarding the reality of media with non--zero \(\chi\)~\cite{lakhtakia1994b,lakhtakia1994c,raab1995,weiglhofer1998}.
        \par
        We can observe that an isotropic chiral medium is a material where there is a linear coupling between one component of the electric (magnetic) polarization, say, \(P_{x}\) (\(M_{x}\)), and the same component of the magnetic (electric) field.  Therefore, for our purpose it is useful to see our initial suspicions confirmed, and to notice is that it is not possible to have a totally spatially isotropic medium, and have such a direct coupling described geometrically by (\ref{consD}) and (\ref{consB2}), for in this case \(\mathbf{g}\) should vanish.  Hence, optical activity in isotropic media cannot be understood in terms of the theory of transformation optics presented in section \ref{riemann}.
\par
        Part of the ambiguity in the form of the constitutive relations for these media comes in deciding whether to use \(\mathbf{B}\) or \(\mathbf{H}\) to describe the coupling of the magnetic field to the electric polarization.  One common alternative form to (\ref{chiral1}--\ref{chiral2}) is the Boys--Post relation~\cite{lakhtakia1994},
        \begin{align}
            \mathbf{D}&=\boldsymbol{\epsilon}_{\text{\tiny{P}}}\cdot\mathbf{E}+\frac{1}{c}\left(\chi_{\text{\tiny{P}}}+i\kappa_{\text{\tiny{P}}}\right)\mathbf{B}\label{chiral3}\\
            \mathbf{H}&=\boldsymbol{\mu}_{\text{\tiny{P}}}^{-1}\cdot\mathbf{B}-\frac{1}{c}\left(\chi_{\text{\tiny{P}}}-i\kappa_{\text{\tiny{P}}}\right)\mathbf{E}\label{chiral4}
        \end{align}
        \par
        In the frequency domain (where these relationships are defined) (\ref{chiral1}--\ref{chiral2}) and (\ref{chiral3}--\ref{chiral4}) are equivalent.  However, it is important that in each case the meaning of the permittivity and permeability is different, as well as the interpretation of the chirality.  Indeed, if we cast (\ref{chiral3}) and (\ref{chiral4}) into the form of (\ref{chiral1}) and (\ref{chiral2}), then we obtain,
        \begin{align*}
            \mathbf{D}&=\left[\boldsymbol{\epsilon}_{\text{\tiny{P}}}+\frac{1}{c^{2}}\left(\chi_{\text{\tiny{P}}}^{2}+\kappa_{\text{\tiny{P}}}^{2}\right)\boldsymbol{\mu}_{\text{\tiny{P}}}\right]\cdot\mathbf{E}+\frac{1}{c}\left(\chi_{\text{\tiny{P}}}+i\kappa_{\text{\tiny{P}}}\right)\boldsymbol{\mu}_{\text{\tiny{P}}}\cdot\mathbf{H}\\
            \mathbf{B}&=\boldsymbol{\mu}_{\text{\tiny{P}}}\cdot\mathbf{H}+\frac{1}{c}\left(\chi_{\text{\tiny{P}}}-i\kappa_{\text{\tiny{P}}}\right)\boldsymbol{\mu}_{\text{\tiny{P}}}\cdot\mathbf{E}
        \end{align*}
        so that we can observe a correspondence in the coupling parameters,
        \begin{equation}
            \begin{tabular}{ccc}
            \(\boldsymbol{\epsilon}_{\text{\tiny{T}}}\)&\(\leftrightarrow\)&\(\boldsymbol{\epsilon}_{\text{\tiny{P}}}+\frac{1}{c^{2}}\left(\chi_{\text{\tiny{P}}}^{2}+\kappa_{\text{\tiny{P}}}^{2}\right)\boldsymbol{\mu}_{\text{\tiny{P}}}\)\\
            \(\boldsymbol{\mu}_{\text{\tiny{T}}}\)&\(\leftrightarrow\)&\(\boldsymbol{\mu}_{\text{\tiny{P}}}\)\\
            \(\chi_{\text{\tiny{T}}}\pm i\kappa_{\text{\tiny{T}}}\)&\(\leftrightarrow\)&\(\boldsymbol{\mu}_{\text{\tiny{P}}}\left(\chi_{\text{\tiny{P}}}\pm i\kappa_{\text{\tiny{P}}}\right)\)
            \end{tabular}
        \end{equation}
        Therefore, in general, when the magnetic susceptibility is anisotropic, whether the chirality is isotropic is relative to the interpretation.  Furthermore, `impedance matching' in the Post constitutive relations does not translate into impedance matching in the Tellegen interpretation.  As noted in~\cite{lakhtakia1994}, true impedance matching must be done relative to the Tellegen parameters.   However, in the limit of small chirality, the difference is negligible.
        \section{Non--Riemannian geometry and isotropic chiral media\label{non-riemann}}
        \par
        From the summary of the theory of transformation optics given in section \ref{riemann}, it is evident that there are two reasons why a Riemannian background has the same effect on the Maxwell equations as a material medium; \begin{inparaenum}[(i)]
        \item the definition of the field tensor in terms of the vector potential is unchanged by the background, so that the co--ordinates may be re--interpreted as if they were Galilean, and;
        \item the expression for the trace of the Christoffel symbols is such that (\ref{final-maxwell-2}) can be written as a total divergence.  Consequently, when the co--ordinate system is re--interpreted as Galilean, the effect of the geometry is that of a charge neutral medium, with no net current passing through any cross section.
        \end{inparaenum}
        \par
        Is it possible to fulfil both these conditions with a more general geometrical background?  Let us explore more general forms of the connection,
        \begin{equation}
            {\Gamma^{\mu}}_{\nu\sigma}=\left\{{}^{\;\mu}_{\nu\sigma}\right\}+{C^{\mu}}_{\nu\sigma}\label{general-connection}
        \end{equation}
        It is instructive to write the non--Riemannian part of the connection, \({C^{\mu}}_{\nu\sigma}\), in terms of component tensors (e.g. see~\cite[Section 3.3]{hurley2000}),
        \[
            {C^{\mu}}_{\nu\sigma}={K^{\mu}}_{\nu\sigma}+{H^{\mu}}_{\nu\sigma},
        \]
        where \({K^{\mu}}_{\nu\sigma}=\frac{1}{2}g^{\alpha\mu}\left[T_{\alpha\nu\sigma}+T_{\nu\alpha\sigma}+T_{\sigma\alpha\nu}\right]\) is the contorsion tensor (recalling that the torsion, \({T^{\mu}}_{\nu\sigma}\) is the anti--symmetric part of the connection), and \({H^{\mu}}_{\nu\sigma}\) represents the non--metricity~\footnote{We should note that the covariant derivative of the Levi--Civita symbol, \(\epsilon_{\mu\nu\sigma\tau}\) vanishes when the trace of the connection equals the trace of the Christoffel symbol~\cite[\textsection{20}]{leonhardt2010}, the contorsion tensor does not alter the trace of the connection.}.  When \({H^{\mu}}_{\nu\sigma}=0\), the connection satisfies the condition, \(\nabla_{\mu}g_{\sigma\tau}=0\), everywhere.  In what follows we assume that \({H^{\mu}}_{\nu\sigma}=0\), so that, \({C^{\mu}}_{\nu\sigma}={K^{\mu}}_{\nu\sigma}\).
	\begin{figure}		
		\includegraphics[width=7cm]{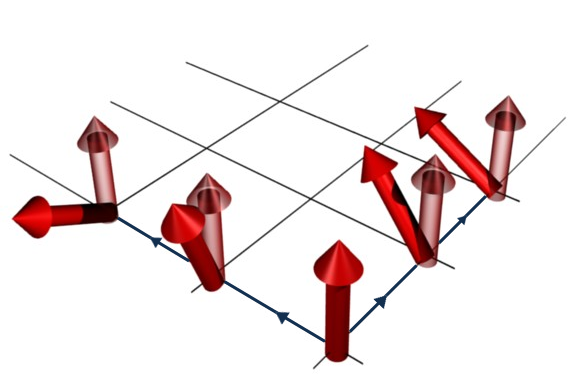}        
		\caption{A completely antisymmetric torsion field, \(T_{\mu\nu\sigma}=-T_{\mu\sigma\nu}=-T_{\nu\mu\sigma}\), does not affect geodesics, and rotates vectors in the plane normal to the direction of parallel transport.  In the figure, parallel transport is being performed along the blue lines.  The transparent vectors show the case when torsion is not included, while the solid vectors show the case when we have a completely antisymmetric torsion field.\label{torsion-figure}}	
	\end{figure}	
	\par
        Almost immediately we hit problems if we add \({K^{\mu}}_{\nu\sigma}\) into the connection.  For if we follow the usual `partial derivative goes to covariant derivative' rule, then the anti--symmetry in the lower indices interferes with the definition of the field tensor, (\ref{final-maxwell-1}),
        \[
            e^{\mu\nu\sigma\tau}\nabla_{\nu}F_{\sigma\tau}=e^{\mu\nu\sigma\tau}\left[\partial_{\nu}F_{\sigma\tau}-{T^{\rho}}_{\sigma\nu}F_{\rho\tau}\right]\stackrel{\text{?}}{=}0.
        \]
        In itself this might not be a problem, did it not break the gauge invariance of the theory.  For if we perform a gauge transformation, \(A^{\prime}_{\mu}=A_{\mu}+\nabla_{\mu}\varphi\), the field tensor ends up depending upon \(\varphi\),
        \begin{align*}
            F_{\mu\nu}^{\prime}&=\nabla_{\mu}A_{\nu}-\nabla_{\nu}A_{\mu}+\left[\nabla_{\mu},\nabla_{\nu}\right]\varphi\\
            &=\nabla_{\mu}A_{\nu}-\nabla_{\nu}A_{\mu}+{T^{\rho}}_{\mu\nu}\nabla_{\rho}\varphi.
        \end{align*}
        Therefore, we reach the conclusion that, with non--zero torsion, the `partial derivative goes to covariant derivative' rule does not produce a gauge invariant theory.  For more extensive coverage of this issue, see~\cite[Section 11.3]{hurley2000} and~\cite{desabbata1994}.  Hence we choose to keep the \emph{definition} of the field tensor the same as in the Riemannian case, \(F_{\mu\nu}=\partial_{\mu}A_{\nu}-\partial_{\nu}A_{\mu}\).
        \par
        The second Maxwell equation does not relate to the definition of the field tensor, and we may suppose that the background geometry modifies this equation with terms including the contorsion tensor.  Applying (\ref{general-connection}) to (\ref{maxwell-free-6}), we obtain,
        \begin{equation}
            \nabla_{\mu}F^{\mu\nu}=\frac{1}{\sqrt{-g}}\partial_{\mu}\left(\sqrt{-g}F^{\mu\nu}\right)+\frac{1}{2}T^{\nu\sigma\tau}F_{\tau\sigma}=0.\label{add-torsion}
        \end{equation}
        In order that this appear as a material medium---c.f. (\ref{maxwell-material-2})---there is a minimal choice for the form of the torsion,
        \begin{equation}
            T^{\nu\sigma\tau}=\mu_{0}\epsilon^{\nu\sigma\tau\mu}\partial_{\mu}\chi\label{torsion-tensor}
        \end{equation}
        where \(\chi\) is an arbitrary single--valued function of \(x^{\mu}\).  The Maxwell equation associated with sources, (\ref{add-torsion}), thus takes the following form,
        \begin{equation}
            \partial_{\mu}\left(\sqrt{-g}F^{\mu\nu}+\frac{1}{2}\chi\mu_{0} e^{\nu\sigma\tau\mu}F_{\tau\sigma}\right)=0\label{second-torsion}
        \end{equation}
        It is interesting that (\ref{second-torsion}) has the same form of coupling to the torsion field as was obtained in~\cite{desabbata1994} through microscopic considerations of the interaction between a space--time torsion field and the vacuum polarization associated with the quantized electromagnetic field (see figure \ref{torsion-figure} for the geometrical interpretation of (\ref{torsion-tensor})).  The divergence-less material field in (\ref{second-torsion}) is,
        \begin{equation}
            G^{\mu\nu}=\frac{\sqrt{-g}}{\mu_{0}}g^{\mu\sigma}g^{\nu\tau}F_{\sigma\tau}+\frac{1}{2}\chi e^{\mu\nu\sigma\tau}F_{\sigma\tau}\label{chiral-geometry}
        \end{equation}
        Due to the appearance of the dual electromagnetic field tensor on the right hand side of (\ref{chiral-geometry}), the presence of the torsion (\ref{torsion-tensor}) in the connection is equivalent to some coupling between like components of the polarization (magnetization) and the magnetic (electric) field.  To be explicit, we write (\ref{chiral-geometry}) in terms of the usual fields, (\(\mathbf{E}\), \(\mathbf{B}\), \(\mathbf{D}\), \(\mathbf{H}\)).
        \par
       Take the simplest case first, and consider an isotropic material, \(g_{\mu\nu}=\text{diag}(1,-n^2,-n^2,-n^2)\).  For this case (\ref{chiral-geometry}) becomes,
        \begin{align}
            \mathbf{D}&=\epsilon_{0}n \mathbf{E}+\frac{\chi}{c} \mathbf{B}\nonumber\\
            \mathbf{H}&=\frac{1}{\mu_{0}n}\mathbf{B}-\frac{\chi}{c} \mathbf{E}\label{tellegen-1}
        \end{align}
        The addition of the torsion, (\ref{torsion-tensor}), into the isotropic geometry may be understood in material terms as an isotropic chiral medium with a non--zero Tellegen parameter, that has equal relative permittivity and permeability as regards the Boys--Post prescription, (\ref{chiral3}--\ref{chiral4}), \(\boldsymbol{\mu}_{\text{\tiny{P}}}/\mu_{0}=\boldsymbol{\epsilon}_{\text{\tiny{P}}}/\epsilon_{0}\).
        \par
        The next simplest case is where the medium is anisotropic, but where the magneto-electric coupling defined by the \(\mathbf{g}\) vector vanishes.  Here, (\ref{chiral-geometry}) gives,
        \begin{align}
            \mathbf{D}&=\boldsymbol{\epsilon}\cdot\mathbf{E}+\frac{\chi}{c}\mathbf{B}\nonumber\\
            \mathbf{H}&=\boldsymbol{\mu}^{-1}\cdot\mathbf{B}-\frac{\chi}{c}\mathbf{E}\label{tellegen-2}
        \end{align}
        where, as before, \(\epsilon_{ij}=\frac{\epsilon_{0}\sqrt{-g}}{g_{00}}\gamma^{-1}_{ij}\) and \(\mu_{ij}=\frac{\mu_{0}\sqrt{-g}}{g_{00}}\gamma^{-1}_{ij}\).  As the torsion, (\ref{torsion-tensor}), is actually a pseudo-tensor, the \(\chi\) parameter changes sign under time reversal or spatial parity inversion: this is consistent with a Tellegen medium, which is not time reversible.  The constitutive relation is somewhat more complicated in the case of a space--time background with a non--vanishing \(\mathbf{g}\) vector as there is an interplay between the magneto-electric coupling due to the metric, coming in the form of the \(\mathbf{g}\) vector, and that coming from the torsion.
        \par
        We conclude that the inhomogeneity of the Tellegen parameter is equivalent to space time torsion in the covariant derivative of (\ref{maxwell-free-6}).  Notice that when the Tellegen parameter is homogeneous, it disappears from the geometry, which again becomes Riemannian: this is consistent with the known invariance of the Maxwell equations under transformations of the fields, where, \(\mathbf{D}\to\mathbf{D}+\eta\mathbf{B}\) and \(\mathbf{H}\to\mathbf{H}-\eta\mathbf{E}\), with uniform \(\eta\) (e.g. see~\cite{lakhtakia1994b,sihvola2008}).
        \par
        It is evident that the more physically important magneto-electric parameter, \(\kappa\), does not arise from the above modification to the space--time connection.  The definition of this quantity, (\ref{chiral3}--\ref{chiral4}), anticipates that it can only arise in the frequency domain, which we now consider.
        \subsection{The frequency domain}
        \par
        In the frequency domain, we consider a purely spatial geometry---\(g_{00}=1\), \(g_{0i}=g_{i0}=0\)---and replace the time derivative with \(-i\omega\).  With this assumption, (\ref{add-torsion}) is,
        \begin{equation}
            \partial_{i}\left(\sqrt{-g}F^{i\nu}\right)+\frac{1}{2}\sqrt{-g}T^{\nu\sigma\mu}F_{\mu\sigma}=i\frac{\omega}{c}\sqrt{-g}F^{0\nu}\label{frequency-chiral}
        \end{equation}
        where we have assumed, as in the previous section, that the additional torsion does not alter the trace of the connection symbol.  One such set of components are as follows,
        \begin{align}
            T^{0ij}&=\frac{i\mu_{0}}{\sqrt{-g}}e^{ijk}\partial_{k}\kappa=-T^{ij0}\nonumber\\
            T^{ijk}&=-\frac{2\mu_{0}}{\sqrt{-g}}e^{ijk}\frac{\omega\kappa}{c}\label{frequency-torsion}
        \end{align}
        Note that, despite appearances, (\ref{frequency-torsion}) transforms as a pseudo-tensor under purely spatial co--ordinate transformations.   We can decompose (\ref{frequency-chiral}) into two equations, one for \(\nu=0\)
        \begin{equation}
            \partial_{i}\left(\frac{\sqrt{-g}}{\mu_{0}}F^{i0}-i\kappa\frac{1}{2}e^{ijk}F_{jk}\right)=0\label{maxwell-frequency-1}
        \end{equation}
        and one for \(\nu=j\),
        \begin{multline}
            \partial_{i}\left(\frac{\sqrt{-g}}{\mu_{0}}F^{ij}-i\kappa e^{ijk}F_{0k}\right)\\
            =i\frac{\omega}{c}\left(\frac{\sqrt{-g}}{\mu_{0}}F^{0j}+i\kappa\frac{1}{2}e^{jik}F_{ik}\right).\label{maxwell-frequency-2}
        \end{multline}
        From the form of (\ref{maxwell-frequency-1}) and (\ref{maxwell-frequency-2}), the torsion given by (\ref{frequency-torsion}) defines the following divergence-less quantity,
        \begin{align*}
            G^{0i}&=\frac{\sqrt{-g}}{\mu_{0}}F^{0i}+i\kappa\frac{1}{2}e^{ijk}F_{jk}\\
            G^{ij}&=\frac{\sqrt{-g}}{\mu_{0}}F^{ij}-i\kappa e^{ijk}F_{0k}
        \end{align*}
       In terms of a Galilean system containing a material medium, these are equivalent to the vector relationships,
        \begin{align*}
            \mathbf{D}&=\boldsymbol{\epsilon}\cdot\mathbf{E}+i\frac{\kappa}{c}\mathbf{B}\\
            \mathbf{H}&=\boldsymbol{\mu}^{-1}\cdot\mathbf{B}+i\frac{\kappa}{c}\mathbf{E},
        \end{align*}
        which define a material medium with a chiral parameter, \(\kappa\), interpreted in the sense of a Boys--Post constitutive relationship.  Combining this result with that of the previous section, a medium with both a chiral parameter and a Tellegen parameter may be defined via a torsion pseudo-tensor with the components,
        \begin{align}
            T^{ijk}&=-2\mu_{0}\epsilon^{ijk}\frac{\omega\kappa}{c}\nonumber\\
            T^{ij0}&=\mu_{0}\epsilon^{ijk}\partial_{k}\left(\chi-i\kappa\right)\nonumber\\
            T^{0ij}&=\mu_{0}\epsilon^{ijk}\partial_{k}\left(\chi+i\kappa\right)\label{frequency-torsion-2}
        \end{align}
        As we are in the frequency domain, we have assumed that the material parameters are independent of time.  Notice that when the material is uniform, the mixed time and space components of the \(T^{\alpha\beta\gamma}\) vanish, and we are left with only the spatial components of the torsion that are proportional to \(\kappa\).  Indeed, the spatial torsion is by far and a way the dominant part of the object, as it is also weighted by the factor, \(\omega/c\).  This is particularly important in the limit of geometrical optics, which we must now investigate.
        \section{Parallel transport and geometrical optics}
        \par
        Transformation optics is a geometrical theory that goes beyond ordinary geometrical optics: as illustrated above, it is concerned with an exact mapping of Maxwell's equations from a geometrical background onto an equivalent material medium.  However, geometrical optics encodes a great deal of the intuitive content of the theory: rays follow geodesics, and polarization is parallel transported~\cite{leonhardt2010,born1999}.  Therefore it is a minimal requirement that the theory of section \ref{non-riemann} bear these intuitions out: geometrical optics should behave as a theory of rays on a background with torsion.  Here we show that this is indeed the case.  The approach of this section is like that of the last, the existing theory is briefly reviewed so that we can then bring out the features of the non--Riemannian modifications.
        \subsection{Riemannian media\label{rmgo}}
        The starting point of geometrical optics is the wave equation, which in a non--chiral, impedance matched medium, can be obtained from the Riemannian form of Maxwell's equations, (\ref{final-maxwell-1}--\ref{final-maxwell-2}).  As noted previously, it is immaterial whether covariant or partial derivatives appear in the definition of the field tensor in a Riemannian geometry.  Therefore the four--dimensional curl of (\ref{final-maxwell-1}) can be written as,
        \begin{equation}
            \epsilon^{\alpha\beta\gamma\mu}\nabla_{\gamma}\left(\epsilon_{\mu\nu\sigma\tau}\nabla^{\nu}F^{\sigma\tau}\right)=0.\label{g1}
        \end{equation}
        As has been assumed throughout, the covariant derivative of the Levi--Civita symbol is zero, and the usual formula, \(\delta^{\alpha\beta\gamma}_{\nu\sigma\tau}=\epsilon^{\alpha\beta\gamma\mu}\epsilon_{\nu\sigma\tau\mu}\)---where \(\delta^{\alpha\beta\gamma}_{\nu\sigma\tau}\) is a \(3\times3\) determinant of Kronecker deltas~\cite{lovelock1975}---can be used to obtain a sum of second derivatives of the field tensor,
        \begin{equation}
             \nabla_{\gamma}\nabla^{\alpha}F^{\beta\gamma}+\nabla_{\gamma}\nabla^{\gamma}F^{\alpha\beta}+\nabla_{\gamma}\nabla^{\beta}F^{\gamma\alpha}=0.\label{levi-civita-expansion}
        \end{equation}
        Applying (\ref{final-maxwell-2}) then gives the wave equation,
        \begin{equation}
             \nabla_{\gamma}\nabla^{\gamma}F_{\alpha\beta}+\left[\nabla^{\gamma},\nabla_{\alpha}\right]F_{\beta\gamma}+\left[\nabla^{\gamma},\nabla_{\beta}\right]F_{\gamma\alpha}=0,\label{wave-equation}
        \end{equation}
        where the commutators of the derivatives are proportional to the contraction of the Riemann curvature tensor, \({R^{\mu}}_{\nu\sigma\tau}\) against each index of the field tensor in turn: e.g. for a four--vector, \(V^{\mu}\), \(\left[\nabla_{\alpha},\nabla_{\beta}\right]V_{\sigma}={R^{\tau}}_{\sigma\beta\alpha}V_{\tau}\).  These curvature terms will not enter the approximation of geometrical optics, because they do not diverge as the wave--length goes to zero.  To see this we write down the field tensor in the form, \(F_{\mu\nu}=f_{\mu\nu}e^{2\pi i S/\lambda}\), where \(\lambda=2\pi c/\omega\), and take the limit of rapidly varying phase: \(\lambda\to0\), which is equivalent to assuming a length scale for the variation of the material properties that is much larger than the wavelength of the optical field.
        \par
        Expanding (\ref{wave-equation}) in powers of \(\lambda^{-1}\), and requiring that the coefficient of each power vanish separately (neglecting the zeroth order as \(\lambda\to0\)), yields the equations of geometrical optics,
        \begin{align}
             \lambda^{-2}:&\;\;\left(\partial^{\alpha}S\right)\left(\partial_{\alpha}S\right)=0\label{gopt1}\\
             \lambda^{-1}:&\;\;\left(\partial_{\gamma} S\right)\nabla^{\gamma}f_{\mu\nu}=\frac{1}{2}f_{\mu\nu}\nabla^{\gamma}\left(\partial_{\gamma} S\right)\label{gopt2}
        \end{align}
        which, as expected, do not include the curvature tensor.  The first of these relations, (\ref{gopt1}) is equivalent to the statement that rays follow geodesics, for if we operate on the left with the covariant derivative, \(\nabla_{\beta}\), and remember that in a Riemannian space--time, the second order derivatives of a scalar commute (the torsion is zero): \(\left[\nabla_{\alpha},\nabla_{\beta}\right]\varphi=0\), then,
        \begin{equation}
            \left(\partial^{\alpha}S\right)\nabla_{\beta}\left(\partial_{\alpha}S\right)=\frac{D^{2}x_{\beta}}{ds^{2}}=0,\label{geodesic}
        \end{equation}
        where the gradient of the phase, \(\partial_{\alpha}S\) is taken to equal the tangent vector to a curve, \(\partial_{\alpha}S=g_{\alpha\beta}dx^{\beta}/ds\), and \(D\equiv dx^{\alpha}\nabla_{\alpha}\).  Equation (\ref{geodesic}) is precisely the rule for the geodesic motion of a material particle in a Riemannian space--time.
        \par
        The second defining equation of geometrical optics, (\ref{gopt2}), illustrates how the polarization changes along a ray.  We write the field tensor amplitude as a bivector that, in the frame we are considering, contains two unit three--vectors, multiplied by an amplitude, \(f_{\mu\nu}=u_{\mu\nu}\mathcal{F}\), where, \(u_{\mu\nu}=(\mathbf{u},\mathbf{v})\): \(u_{i}u^{i}=1\); \(v_{i}v^{i}=1\); \(u_{i}k^{i}=v_{i}k^{i}=u_{i}v^{i}=0\).  The tensor, \(u_{\mu\nu}\) indicates the direction of the polarization, and (\ref{gopt2}) becomes,
        \begin{equation}
            \frac{1}{2}\nabla_{\gamma}\left(\mathcal{F}^{2}\frac{dx^{\gamma}}{ds}\right)u_{\mu\nu}+\mathcal{F}^{2}\left(\frac{Du_{\mu\nu}}{ds}\right)=0.\label{polarization-equation}
        \end{equation}
        Both terms to the left of the equality in (\ref{polarization-equation}) must vanish separately if the unit polarization vectors defined within \(u_{\mu\nu}\) are to remain unit vectors at all points along a ray (i.e. otherwise they would grow or diminish exponentially with the relative change in the energy density).  Consequently,
        \begin{align}
             \nabla_{\gamma}\left(\mathcal{F}^{2}\frac{dx^{\gamma}}{ds}\right)&=0\label{continuity}\\
             \frac{Du_{\mu\nu}}{ds}&=0.\label{parallel-transport}
        \end{align}
        Explicitly, (\ref{continuity}) is equivalent to the continuity of the energy momentum along the ray, and (\ref{parallel-transport}) to the statement that the direction of the polarization is parallel transported along the ray.  These are the essential results of geometrical optics in a Riemannian medium.
        \subsection{non--Riemannian media\label{torsion-go}}
        \par
        If we allow for the possibility of a space time with torsion as discussed above, then it is not immediately obvious whether the equations of geometrical optics will still carry the simple structure where rays follow geodesics, and polarization is parallel transported: although the Maxwell equations remain identical in form to (\ref{final-maxwell-1}) and (\ref{final-maxwell-2}) throughout section \ref{non-riemann}, this was achieved through imposing the definition of the field tensor, and we should check that this imposition has not spoilt the geometrical interpretation.   We shall show that, due to the particular form of the torsion, (\ref{frequency-torsion-2}), the geometrical optics limit is unaffected; rays still follow geodesics, and the polarization is still parallel transported.
        \par
        The analogous situation to that represented by equation (\ref{g1}) contains an additional contribution that arises because of the use of partial, rather than covariant derivatives to define the field tensor,
        \[
           \epsilon^{\alpha\beta\gamma\mu}\nabla_{\gamma}\epsilon_{\mu\nu\sigma\tau}\left(\nabla^{\nu}F^{\sigma\tau}-{T_{\alpha}}^{\nu\sigma}F^{\alpha\tau}\right)=0
        \]
        Following the same procedure that led to (\ref{levi-civita-expansion}), and expanding the contraction of the Levi--Civita symbols,
        \begin{multline}
            \nabla_{\gamma}\nabla^{\gamma}F^{\alpha\beta}+\left[\nabla_{\gamma},\nabla^{\alpha}\right]F^{\beta\gamma}+\left[\nabla_{\gamma},\nabla^{\beta}\right]F^{\gamma\alpha}\\
            \nabla_{\gamma}\left({T_{\rho}}^{\alpha\gamma}F^{\rho\beta}+{T_{\rho}}^{\beta\alpha}F^{\rho\gamma}+{T_{\rho}}^{\gamma\beta}F^{\rho\alpha}\right)=0,\label{chiral-field-derivative}
        \end{multline}
        it is clear that there are terms in addition to the wave operator, \(\nabla_{\gamma}\nabla^{\gamma}F^{\alpha\beta}\), that involve not just the contraction of geometric quantities against the field tensor, but also against \emph{derivatives} of the field tensor.  Such terms will remain in the equations of geometrical optics, and we must investigate them further.
        \par
        The commutator of the derivatives of the field tensor, \(\left[\nabla_{\gamma},\nabla^{\alpha}\right]F^{\beta\gamma}\), now also contains terms involving the torsion tensor,
        \begin{equation}
             \left[\nabla_{\alpha},\nabla_{\beta}\right]F^{\mu\nu}={R^{\mu}}_{\sigma\alpha\beta}F^{\sigma\nu}+{R^{\nu}}_{\sigma\alpha\beta}F^{\mu\sigma}+{T^{\rho}}_{\alpha\beta}\nabla_{\rho}F^{\mu\nu}.\label{torsion-commutator}
        \end{equation}
        Using (\ref{torsion-commutator}) and applying (\ref{maxwell-free-4}), we can group the terms in addition to the wave operator in (\ref{chiral-field-derivative}).  The zeroth order terms involving no derivatives of the fields are found to be,
        \begin{multline*}
             0^{\text{th}}:\;\;g^{\alpha\sigma}\left({R^{\beta}}_{\rho\gamma\sigma}F^{\rho\gamma}+{R^{\gamma}}_{\rho\gamma\sigma}F^{\beta\rho}\right)\\
             -g^{\beta\sigma}\left({R^{\alpha}}_{\rho\gamma\sigma}F^{\rho\gamma}+{R^{\gamma}}_{\rho\gamma\sigma}F^{\alpha\rho}\right)\\
             +F^{\rho\beta}\nabla_{\gamma}{T_{\rho}}^{\alpha\gamma}-F^{\rho\alpha}\nabla_{\gamma}{T_{\rho}}^{\gamma\beta}+F^{\rho\gamma}\nabla_{\gamma}{T_{\rho}}^{\beta\alpha}
        \end{multline*}
        meanwhile, the terms that are first order in the derivatives of the field tensor are,
        \begin{multline}
             1^{\text{st}}:\;\;\left(T^{\rho\alpha\gamma}-T^{\gamma\rho\alpha}\right)\nabla_{\gamma}{F_{\rho}}^{\beta}-\left(T^{\rho\beta\gamma}-T^{\gamma\rho\beta}\right)\nabla_{\gamma}{F_{\rho}}^{\alpha}.\label{first-order}
        \end{multline}
        For a general form of \(T^{\alpha\beta\gamma}\), (\ref{first-order}) is non--zero, and a simple geometric description will not apply.  However it is immediately clear that the description of Tellegen media given by (\ref{torsion-tensor}) makes these first order terms vanish.
        \par
        In treating (\ref{first-order}) in the frequency domain, we assume, as in~\cite{lindell1989}, that the optical activity of the chiral parameter, \(\kappa\) is such that the polarization is only slightly changed over each optical cycle: i.e. that \(\omega\kappa/c\) does not diverge as \(\lambda\to0\).  This means that the quantity \(\kappa\) is of order \(\lambda\), and the non--zero part of (\ref{frequency-torsion-2}) in the limit equals only the spatial part of the torsion, \(T^{ijk}\) (not including the Tellegen parameter, the contribution of which we have shown to equal zero), so that (\ref{first-order}) also vanishes in this case.  This proves that the limit of geometrical optics involves only the wave operator and zeroth order terms, just as in the Riemannian case.

        \par
        We have therefore established that, as in the usual situation presented in section \ref{rmgo}, only the wave operator matters in the limit of geometrical optics,
        \[
            \nabla_{\gamma}\nabla^{\gamma}F^{\mu\nu}=0.
        \]
       For slowly varying torsion and small curvature and torsion in comparison to \(1/\lambda\) this is the equation obeyed by the exact solution to Maxwell's equations, to a good approximation.  The geometrical understanding of the theory becomes more complicated when the torsion is rapidly varying, just as it does when the curvature is large in the usual theory of transformation optics.  So our formalism passes the first test.
       \par
       Inserting the ansatz for the field tensor as in the previous section, we have, again (\ref{gopt1}) and (\ref{gopt2}).  However, the meaning of these equations is now slightly different.  For if we take (\ref{gopt1}) and attempt to derive the geodesic equation as before then we find an additional term,
        \begin{align*}
            \left(\partial_{\alpha}S\right)\nabla_{\beta}\left(\partial^{\alpha}S\right)&=\left(\partial_{\alpha}S\right)\nabla^{\alpha}\left(\partial_{\beta}S\right)+\left(\partial_{\alpha}S\right)\left[\nabla_{\beta},\nabla^{\alpha}\right]S\\
            &=\left(\partial_{\alpha}S\right)\nabla^{\alpha}\left(\partial_{\beta}S\right)+T_{\sigma\beta\gamma}(\nabla^{\sigma}S)(\nabla^{\gamma}S)
        \end{align*}
        For the same reason that this limit works in the first place, namely the vanishing of the terms in (\ref{first-order}), this additional contribution vanishes and we have,
        \begin{equation}
            \frac{D^{2}x^{\beta}}{ds^{2}}=\frac{d^2x^{\beta}}{ds^2}+{\Gamma^{\beta}}_{\sigma\alpha}\frac{dx^{\sigma}}{ds}\frac{dx^{\alpha}}{ds}=0\label{autoparallel}
        \end{equation}
        This is the equation for an auto-parallel rather than a geodesic~\cite[Section 10]{kleinert2006}, as it contains the full connection and not only the Christoffel symbol: this equation formally determines the straightest line between two points and not the shortest.  However, again due to the antisymmetry of (\ref{frequency-torsion-2}) in all indices, the contribution of the contorsion is zero (i.e. chiral media are equivalent to a geometry with a \({K^{\mu}}_{\nu\sigma}\) that is antisymmetric in the lower two indices), and rays follow geodesics: \({\Gamma^{\beta}}_{\sigma\alpha}\to\left\{{}^{\;\beta}_{\sigma\alpha}\right\}\).
        \par
        The derivation of the equivalent of (\ref{polarization-equation}) is unaltered in this situation, and so (\ref{continuity}) and (\ref{parallel-transport}) remain in the same form.  Firstly the propagation of energy--momentum,
        \[
             \nabla_{\gamma}\left(\mathcal{F}^{2}\frac{dx^{\gamma}}{ds}\right)=\partial_{\gamma}\left(\mathcal{F}^{2}\frac{dx^{\gamma}}{ds}\right)+\mathcal{F}^{2}{\Gamma^{\gamma}}_{\alpha\gamma}\frac{dx^{\alpha}}{ds}=0.
        \]
        The contorsion that gave rise to the chirality did not alter the trace of the connection.  Therefore energy--momentum propagates relative to geodesics, as in the Riemannian case: this is consistent with the equivalence of (\ref{autoparallel}) to geodesic motion.  Meanwhile, the propagation of the polarization along the ray \emph{is} affected by the presence of the torsion,
        \begin{equation}
             \frac{Du_{\mu\nu}}{ds}=\frac{d u_{\mu\nu}}{ds}-{\Gamma^{\sigma}}_{\mu\alpha}u_{\sigma\nu}\frac{dx^{\alpha}}{ds}-{\Gamma^{\sigma}}_{\nu\alpha}u_{\mu\sigma}\frac{dx^{\alpha}}{ds}=0.\label{polarization-transport}
        \end{equation}
        So the formalism appears to be consistent with the idea that weak chirality should act only to rotate polarization during propagation (c.f. figure \ref{torsion-figure}).
        \par
        \emph{In summary}:  for a chiral medium that rotates the polarization by a finite amount over a typical length scale, and where the change in the chiral parameter, \(\kappa\), is not significant over a wave--length, the geometrical optics of chiral media requires that we add torsion into the connection.  The form of the torsion, (\ref{frequency-torsion-2}), is such that the propagation of a ray is unaffected---geodesics are equivalent to auto-parallels---while the parallel transport of the polarization is modified.  This result also holds for Tellegen media, where \(\chi\) does not vary too rapidly.
%
%
        \section{Applications}
%
%
        \subsection{A homogeneous, isotropic, chiral medium}
        \par
        The simplest test of this theory is to apply it to the simplest kind of chiral medium; one that is homogenous.  Note that throughout this section and the next we implicitly work in the frequency domain, despite using four dimensional notation.  For an isotropic, homogeneous medium, the theory of sections \ref{riemann} and \ref{non-riemann} prescribe that the metric and torsion should be given by,
        \[
            g_{\mu\nu}=\left(\begin{matrix}
                     1&0&0&0\\
                     0&-n^2&0&0\\
                     0&0&-n^2&0\\
                     0&0&0&-n^2
                     \end{matrix}\right),
        \]
        and,
        \[
            T^{0ij}=T^{i0j}=0;\;\;T^{ijk}=-2\mu_{0}\frac{\omega\kappa}{c}\epsilon^{ijk}
        \]
        Due to the assumed uniformity of \(n\), the Christoffel symbols vanish, and the connection is equal to the contorsion tensor,
        \[
            {\Gamma^{\mu}}_{\nu\sigma}=\frac{1}{2}g^{\mu\alpha}\left[T_{\nu\alpha\sigma}+T_{\sigma\alpha\nu}+T_{\alpha\nu\sigma}\right]=\frac{1}{2}g^{\mu\alpha}T_{\alpha\nu\sigma}
        \]
        Therefore (\ref{autoparallel}) becomes,
        \begin{equation}
            \frac{d^{2}x^{\beta}}{ds^{2}}=0.\label{straight-line}
        \end{equation}
        Rays follow straight lines.  Yet, the polarization is changed along each ray, as is clear from (\ref{polarization-transport}).  For instance if we take the unit vector for the electric field, \(u_{0i}=u_{i}\), then it changes according to,
        \begin{equation}
            \frac{d u_{i}}{ds}=-\frac{\alpha}{n}e_{ijk}k_{j}u_{k},\label{change-of-polarization}
        \end{equation}
        where \(\alpha=\mu_{0}\omega\kappa/c\) and \(k_{j}=dx_{j}/ds\).  Suppose that a ray travels along the \(x\)--axis, \(\mathbf{k}=(n,0,0)\).  In this case, \(\mathbf{u}=(0,u_{y},u_{z})\), and (\ref{change-of-polarization}) yields two coupled equations,
        \begin{align}
            \frac{du_{y}}{ds}&=\alpha u_{z}\nonumber\\
            \frac{du_{z}}{ds}&=-\alpha u_{y}\label{unit-vector-change}
        \end{align}
        From the definition of \(dx^{i}/ds\), \(dx^{i}dx_{i}=ds^{2}\), the line element on the ray is, \(ds=n dx\). Therefore the covariant unit vector, \(\mathbf{u}\) has the following form,
        \begin{equation}
            \mathbf{u}=n\left(0,\sin{(n\alpha x)},\cos{(n\alpha x)}\right).\label{polarization-change-1}
        \end{equation}
        An almost identical calculation for the unit vector of the magnetic field, \(\mathbf{v}\) shows that \(\mathbf{v}\) also satisfies (\ref{unit-vector-change}).  Applying the definition of the field tensor,
        \begin{equation}
            \mathbf{v}=n\left(0,-\cos{(n\alpha x)},\sin{(n\alpha x)}\right)\label{polarization-change-2}
        \end{equation}
        From (\ref{straight-line}), (\ref{polarization-change-1}), and (\ref{polarization-change-2}), the field in the material is proportional to,
        \begin{align}
            \mathbf{E}&=\left(0,\sin{(n\alpha x)},\cos{(n\alpha x)}\right)e^{i\frac{\omega}{c}\left(nx-ct\right)}\label{E-chiral}\\
            \mathbf{B}&=\frac{n}{c}\left(0,-\cos{(n\alpha x)},\sin{(n\alpha x)}\right)e^{i\frac{\omega}{c}\left(nx-ct\right)}\label{B-chiral}
        \end{align}
        In the weakly chiral limit in which we are working, (\ref{E-chiral}) and (\ref{B-chiral}) are the solutions to the Maxwell equations.  To see this we consider the wave equation that arises from the usual Maxwell equations (\(\mathbf{\mu}/\mu_{0}=\mathbf{\epsilon}/\epsilon_{0}\)) with the constitutive relations (\ref{chiral3}) and (\ref{chiral4}),
        \begin{equation}
             \boldsymbol{\nabla}^{2}\mathbf{E}+\frac{2\mu_{0}n\kappa\omega}{c}\boldsymbol{\nabla}\times\mathbf{E}+\frac{n^{2}\omega^{2}}{c^{2}}\mathbf{E}=0.\label{maxwell-check}
        \end{equation}
        Substituting in an electric field of the form \(\mathbf{u}(x)\,e^{i n\frac{\omega}{c}x}\) into (\ref{maxwell-check}) gives,
        \begin{equation}
             \frac{\partial\mathbf{u}}{\partial x}+\frac{\mu_{0}n\omega\kappa}{c}\hat{\mathbf{x}}\times\mathbf{u}=\frac{ic}{2n\omega}\left[\frac{\partial^{2}\mathbf{u}}{\partial x^{2}}+\frac{2\mu_{0}n\kappa\omega}{c}\boldsymbol{\nabla}\times\mathbf{u}\right]\label{from-maxwell}
        \end{equation}
        The right hand side of (\ref{from-maxwell}) is proportional to \(\lambda\) times a quantity of order unity.  Therefore, in the approximation of (\ref{E-chiral}--\ref{B-chiral}),
        \[
             \frac{\partial\mathbf{u}}{\partial x}+\frac{\mu_{0}n\omega\kappa}{c}\hat{\mathbf{x}}\times\mathbf{u}=0,
        \]
        which is identical to (\ref{unit-vector-change}).  This proves that in the case of an isotropic, homogeneous, chiral medium, with \(\mu_{0}\kappa\ll1\), non--Riemannian geometrical optics is equivalent to the solution of Maxwell's equations. 
%
%
        \subsection{Maxwell's fish eye lens}
        \par
        In the formalism of transformation optics presented in section~\ref{riemann}, geometry is implemented for the purpose of directing rays, and polarization is a bystander, and must respond in a way that is determined by the geodesics.  However, one may wish to maintain a given polarization throughout a device, or change it in some prespecified manner.  Here we show, using the simplest example of a curved geometry for light---the Maxwell fish eye~\cite{maxwell1854,luneburg1964,born1999,leonhardt2010}---that torsion can be used to control the polarization of light without affecting the geodesics.  We should note that chiral media have previously been considered for `correcting' polarization in a variant of the planar fish eye~\cite{lindell1989}.
	\begin{figure}
		(a)\includegraphics[width=7cm]{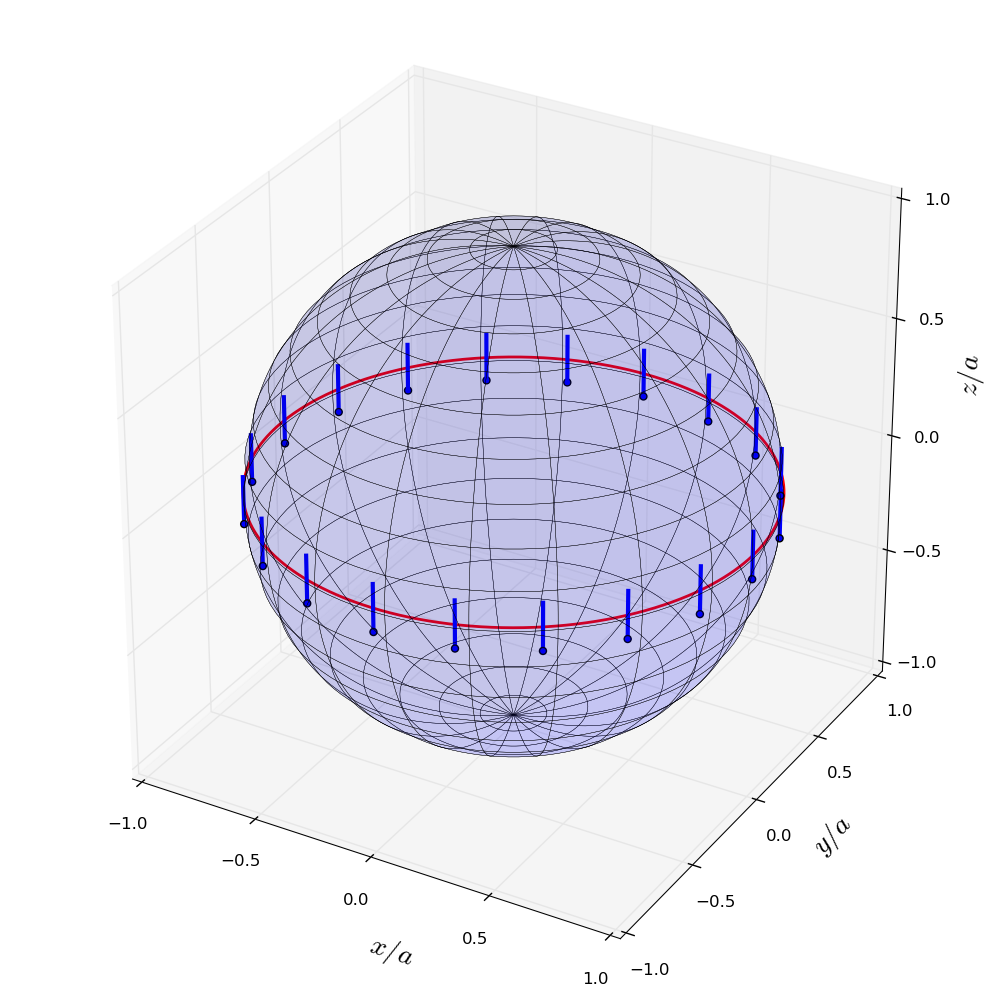}\\
		(b)\includegraphics[width=7cm]{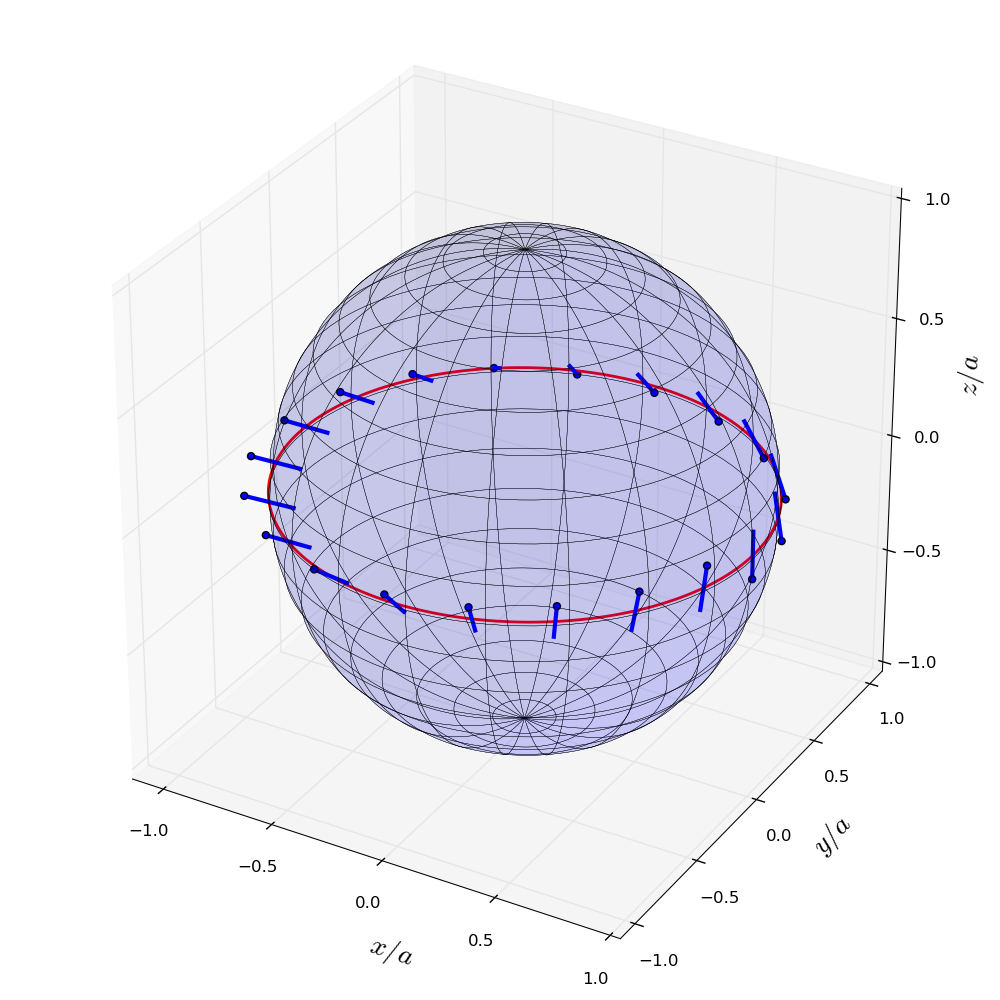}
		\caption{In Maxwell's fish eye lens, the ray trajectories are the great circles of a 3--sphere, as illustrated for the special case of \(r=a\) in (a) and (b).  (a) shows the usual situation, where the refractive index determines both the ray trajectory (red line) and the behaviour of the polarization (blue lines).  Meanwhile (b) shows that transformation optics with the inclusion of the geometrical torsion, (\ref{maxwell-torsion}) (i.e. isotropic chirality), allows for the manipulation of polarization without affecting the geodesics.\label{transport-figure}}	
	\end{figure}
        \par
        The fish eye is a continuous medium in which the behaviour of light maps onto the free motion on the surface of a sphere (this may be a 2--sphere or a 3--sphere, depending upon whether the medium is planar, or truly three dimensional).  It has recently recieved much attention, in part due to its ability to periodically perfectly reconstruct an initial optical pulse as it propagates through the medium, as well as sub--diffraction resolution~\cite{leonhardt2009b,leonhardt2010b,benitez2010,minano2010}.
	\par	
	If the behaviour of light in terms of lab co--ordinates \((x,y,z)\) corresponds to motion on the surface of a 3--sphere (a hypersphere), the optical line element can be written in a form corresponding to an isotropic medium,
        \begin{align}
            dl^{2}&=n(r)^{2}\left(dx^{2}+dy^{2}+dz^{2}\right)\nonumber\\
            &=a^{2}\left[d\Theta^{2}+\sin^{2}(\Theta)\left(d\Phi^{2}+\sin^{2}(\Phi)d\chi^{2}\right)\right],\label{maxwell-line-element}
        \end{align}
        where \(r=\sqrt{x^{2}+y^{2}+z^{2}}\), and, \(a\), (\(\Theta,\Phi,\chi\)), are the radius and angular co--ordinates of the equivalent 3--sphere.
        \par
        A spherical geometry has no boundary, whereas physical space has a `boundary' at infinity.  Therefore, as \(r\to\infty\) in physical space, \(n(r)\) should be such that the radius of any circle surrounding the origin, as experienced by a light ray, \(2\pi n(r) r\), goes to zero.   Also, the length of any optical path, \(\int_{0}^{\infty}n(r)dr\), should equal \(\pi a\), where \(a\) is the radius of the sphere.  A refractive index fulfilling both these requirements is,
        \begin{equation}
             n(r)=2a\frac{d}{dr}\arctan{(r/a)}=\frac{2}{1+(r/a)^{2}}.\label{maxwell-n}
        \end{equation}
        If we introduce spherical polar co--ordinates in the lab system---\(dx^{2}+dy^{2}+dz^{2}=dr^{2}+r^{2}(d\theta^{2}+\sin^{2}(\theta)d\phi^{2})\) in (\ref{maxwell-line-element})---then the observation that the angles, \(\theta,\phi\), must, by symmetry, equal the corresponding angles on the 3--sphere, \(\theta,\phi=\Phi,\chi\) leads, via (\ref{maxwell-n}), to the identification: \(r=a\tan(\Theta/2)\).  Performing this transformation of the radial co--ordinate in the first line of (\ref{maxwell-line-element}) gives the line element on the second line of (\ref{maxwell-line-element}), justifying (\ref{maxwell-n}).
        \par
        The metric tensor associated with Maxwell's fish eye can be immediately written down from (\ref{maxwell-line-element}),
	\[
		g_{\mu\nu}=\left(\begin{matrix}
		1&0&0&0\\
		0&-a^2&0&0\\
		0&0&-a^2\sin^{2}(\Theta)&0\\
		0&0&0&-a^2\sin^{2}(\Theta)\sin^{2}(\Phi)\\
		\end{matrix}\right)
	\]
	This corresponds to the non--zero Christoffel symbols; \(\left\{{}^{\,1}_{22}\right\}=-\sin(\Theta)\cos(\Theta)\); \(\left\{{}^{\,1}_{33}\right\}=-\sin(\Theta)\cos(\Theta)\sin^{2}(\Phi)\); \(\left\{{}^{\,2}_{12}\right\}=\cot(\Theta)\); \(\left\{{}^{\,2}_{33}\right\}=-\sin(\Phi)\cos(\Phi)\); \(\left\{{}^{\,3}_{13}\right\}=\cot(\Theta)\); and \(\left\{{}^{\,3}_{23}\right\}=\cot(\Phi)\).  These symbols determine the trajectories of the light rays, as well as the change in the polarization during propagation.  In particular, light rays with motion along the \(\Phi,\chi\) axes are confined to a 2--sphere (of constant \(\Theta\)) when \(\Theta=\pi/2\) (\(r=a\) in the lab system).  As a visual example, let's examine the motion of light rays and transport of polarization in this specific case (see figure~\ref{transport-figure}).
	\par	
	The geodesic equations (\ref{geodesic}) on the 2--sphere defined by \(\Theta=\pi/2\) are,
	\begin{align}
		\frac{d^{2}\Theta}{ds^{2}}&=0\nonumber\\
		\frac{d^{2}\Phi}{ds^{2}}&=\sin{(\Phi)}\cos{(\Phi)}\frac{d\chi}{ds}\frac{d\chi}{ds}\nonumber\\
		\frac{d^{2}\chi}{ds^{2}}&=-2\cot{(\Phi)}\frac{d\Phi}{ds}\frac{d\chi}{ds},\label{geodesic-maxwell}	
	\end{align}
	so the rays remain confined on the 2--sphere.  The change of the polarization is similarly obtained, through applying (\ref{parallel-transport}),
	\begin{align}
		\frac{du_{01}}{ds}&=0\nonumber\\
		\frac{du_{02}}{ds}&=\cot{(\Phi)}u_{03}\frac{d\chi}{ds}\nonumber\\
		\frac{du_{03}}{ds}&=\cot{(\Phi)}u_{03}\frac{d\Phi}{ds}-\sin{(\Phi)}\cos{(\Phi)}u_{02}\frac{d\chi}{ds}\label{polarization-maxwell}
	\end{align}
	Equations (\ref{geodesic-maxwell}) can be integrated to give the ray trajectory,
	\[
		\frac{d\Phi}{d\chi}=\pm\sqrt{\sin^{2}(\Phi)(l/l_z)^{2}-1},
	\]
	where \(l\) and \(l_z\) are the constants of integration in (\ref{geodesic-maxwell}), and the sign of the derivative changes when the square root goes to zero.  Noting the symmetry of the medium, and setting \(\Phi=\pi/2\), \(l=l_z\), it is clear that rays propagate along the great circles and polarization has a constant orientation along each ray (figure \ref{transport-figure} a).  In Riemannian transformation optics, we can only manipulate the propagation of polarization along a ray through changing the geodesics.  However, the additional geometrical property of torsion, outlined in sections \ref{non-riemann} and \ref{torsion-go} allows us to `twist' the co--ordinate lines on the sphere so that the polarization is changed, while leaving the geodesics unaltered.
	\par	
	For instance, if we wished to generate TE polarization at one point on the sphere (e.g. \(\Theta=\pi/2,\Phi=\pi/2,\chi=0\)), and have it arrive at the antipode (\(\Theta=\pi/2,\Phi=\pi/2,\chi=\pi\)) with TM polarization, then we could add the following torsion into the connection (this picture is accurate so long as \(a\gg\lambda\)),
	\begin{equation}
		T_{ijk}=a^{2}\epsilon_{ijk}\label{maxwell-torsion}
	\end{equation}
	The behaviour of the polarization on the surface of the 2--sphere with the addition of (\ref{maxwell-torsion}) is shown in figure \ref{transport-figure}.\\
        \section{Conclusions}
        We have shown that non--Riemannian geometry can be introduced into the Maxwell equations to describe inhomogeneous media with isotropic chirality.  If the definition of the field tensor is kept in the same gauge invariant form as in a Riemannian geometry, then we have shown that a background with a particular kind of torsion can reproduce the correct constitutive relations for such media.  Furthermore, if the chirality produces only a small amount of optical activity over a single optical cycle, and varies by only a small amount over a wavelength, then we have also shown that geometrical optics behaves exactly as if on a background with a non--zero torsion.  This formalism allows for a greater degree of control over polarization within the theory of transformation optics, and we have given an example of how torsion can be implemented to the change of polarization without changing the geodesics.\\
        \acknowledgments
        This research is supported by an EPSRC postdoctoral fellowship award.  I thank Tom Philbin and Ulf Leonhardt for useful discussions, suggestions and corrections (and patience).
        \bibliography{refs}

\begin{thebibliography}{10}

\bibitem{leonhardt2006b}
U.~Leonhardt and T.~G. Philbin.
\newblock {\em New J. Phys.}, 8:247, 2006.

\bibitem{volume2}
L.~D. Landau and E.~M. Lifshitz.
\newblock {\em The Classical Theory of Fields}.
\newblock Butterworth-Heinemann, Oxford, 2003.

\bibitem{leonhardt2010}
U.~Leonhardt and T.~G. Philbin.
\newblock {\em Geometry and Light: The Science of Invisibility}.
\newblock Dover, New York, 2010.

\bibitem{chen2010}
H.~Chen, C.~T. Chan, and S.~Ping.
\newblock {\em Nat. Mat.}, 9:387, 2010.

\bibitem{leonhardt2006}
U.~Leonhardt.
\newblock {\em Science}, 312:1777, 2006.

\bibitem{pendry2006}
J.~B. Pendry, D.~Schuring, and D.~R. Smith.
\newblock {\em Science}, 312:1780, 2006.

\bibitem{shalaev2010}
W.~Cai and V.~Shalaev.
\newblock {\em Optical Metamaterials}.
\newblock Springer, London, 2010.

\bibitem{philbin2008}
T.~G. Philbin, C.~Kuklewicz, S.~Robertson, S.~Hill, F.~K{\"o}nig, and
  U.~Leonhardt.
\newblock {\em Science}, 319:1367, 2008.

\bibitem{belgiorno2010}
F.~Belgiorno, S.~L. Cacciatori, M.~Clerici, V.~Gorini, G.~Ortenzi, L.~Rizzi,
  E.~Rubino, V.~G. Sala, and D.~Faccio.
\newblock {\em Phys. Rev. Lett.}, 105:203901, 2010.

\bibitem{born1999}
M.~Born and E.~Wolf.
\newblock {\em Principles of Optics}.
\newblock Cambridge University Press, 1999.

\bibitem{leonhardt2009}
U.~Leonhardt and T.~Tyc.
\newblock {\em Science}, 323:110, 2009.

\bibitem{pendry2000}
J.~B. Pendry.
\newblock {\em Phys. Rev. Lett.}, 85:3966, 2000.

\bibitem{leonhardt2009b}
U.~Leonhardt.
\newblock {\em New J. Phys.}, 11:093040, 2009.

\bibitem{Note1}
The conformal invariance of Maxwell's equations (i.e. invariance under \(g_{\mu
  \nu }\to f(x^\sigma )g_{\mu \nu }\)) has the consequence that out of \(n\)
  independent space--time metric components there are \(n-1\) independent
  material parameters.

\bibitem{Note2}
When considering complex entries in the \(\epsilon \), \(\mu \), and
  electric--magnetic coupling tensors, the number of under-described parameters
  is larger.

\bibitem{andrade2004}
L.~C. Garcia~de Andrade.
\newblock {\em Phys. Rev. D}, 70:064004, 2004.

\bibitem{andrade2005}
L.~C. Garcia~de Andrade.
\newblock {\em Phys. Lett. A}, 346:327--329, 2005.

\bibitem{ward1996}
A.~J. Ward and J.~B. Pendry.
\newblock {\em J. Mod. Opt.}, 43:773, 1996.

\bibitem{hurley2000}
D.~J. Hurley and M.~A. Vandyck.
\newblock {\em Geometry, Spinors and Applications}.
\newblock Springer, Praxis, Chichester, 2000.

\bibitem{lovelock1975}
D.~Lovelock and H.~Rund.
\newblock {\em Tensors, Differential Forms and Variational Principles}.
\newblock Dover, New York, 1975.

\bibitem{Note3}
Throughout we use the term Galilean in accordance with Landau and Lifshitz,
  i.e. to mean the system of co--ordinates where the metric equals, \(\eta
  _{\mu \nu }=\protect \text {diag}(1,-1,-1,-1)\), everywhere.

\bibitem{Note4}
The minus sign is introduced into the square root of \(g\) for the sake of
  convention: physical space--time has the signature \((1,3)\), and therefore a
  negative value for the determinant, \(g\). However, transformation optics is
  free to explore media that are equivalent to space--times with arbitrary
  signature (e.g. see~\cite {smolyaninov2010}).

\bibitem{volume8}
L.~D. Landau, E.~M. Lifshitz, and L.~P. Pitaevskii.
\newblock {\em The Electrodynamics of Continuous Media}.
\newblock Butterworth-Heinemann, Oxford, 2004.

\bibitem{lakhtakia1994}
A.~Lakhtakia.
\newblock {\em Beltrami Fields in Chiral Media}.
\newblock World Scientific, Singapore, 1994.

\bibitem{lindell1994}
I.~V. Lindell, A.~H. Sihvola, S.~A. Tretyakov, and A.~J. Viitanen.
\newblock {\em Electromagnetic waves in Chiral and Bi--anisotropic Media}.
\newblock Routledge, 1994.

\bibitem{Note5}
The Drude--Born--Federov constitutive relations modify the Drude--Born
  relations to describe media with symmetric magneto-electric coupling (see
  e.g.~\cite [\textsection {3-2}]{lakhtakia1994}), coupling \(\protect \mathbf
  {D}\) to \(\protect \boldsymbol {\nabla }\times \protect \mathbf {E}\) as
  well as \(\protect \mathbf {B}\) to \(\protect \boldsymbol {\nabla }\times
  \protect \mathbf {H}\). Clearly, these relationships are also non--local.

\bibitem{tellegen1948}
B.~D.~H. Tellegen.
\newblock {\em Philips Res. Rept.}, 3:81, 1948.

\bibitem{lakhtakia1994b}
A.~Lakhtakia.
\newblock {\em Int. J. Infrared and Millimeter Waves}, 15:1625, 1994.

\bibitem{lakhtakia1994c}
A.~Lakhtakia and W.~S. Weiglhofer.
\newblock {\em IEEE Trans. Microwave Theory and Techniques}, 42:1715, 1994.

\bibitem{raab1995}
R.~E. Raab and A.~H. Sihvola.
\newblock {\em J. Phys. A: Math. Gen.}, 30:1335, 1997.

\bibitem{weiglhofer1998}
W.~S. Weiglhofer and A.~Lakhtakia.
\newblock {\em J. Phys. A: Math. Gen.}, 31:1113, 1998.

\bibitem{Note6}
We should note that the covariant derivative of the Levi--Civita symbol,
  \(\epsilon _{\mu \nu \sigma \tau }\) vanishes when the trace of the
  connection equals the trace of the Christoffel symbol~\cite [\textsection
  {20}]{leonhardt2010}, the contorsion tensor does not alter the trace of the
  connection.

\bibitem{desabbata1994}
V.~de~Sabbata and C.~Sivaram.
\newblock {\em Spin and Torsion in Gravitation}.
\newblock World Scientific, Singapore, 1994.

\bibitem{sihvola2008}
A.~Sihvola and S.~Tretyakov.
\newblock {\em Optik}, 119:247, 2008.

\bibitem{lindell1989}
I.~V. Lindell, A.~H. Sihvola, A.~J. Viitanen, and S.~A. Tretyakov.
\newblock {\em Proc. 19th European Microwave Conference}, page 534, 1989.

\bibitem{kleinert2006}
H.~Kleinert.
\newblock {\em Path Integrals in Quantum Mechanics, Statistics, Polymer Physics
  and Financial Markets}.
\newblock World Scientific, Singapore, 2006.

\bibitem{maxwell1854}
J.~C. Maxwell.
\newblock {\em Camb. Dublin Math J.}, 8:188, 1854.

\bibitem{luneburg1964}
R.~K. Luneburg.
\newblock {\em Mathematical Theory of Optics}.
\newblock University of California Press, Berkeley, CA, 1964.

\bibitem{leonhardt2010b}
U.~Leonhardt and T.~G. Philbin.
\newblock {\em Phys. Rev. A}, 81:011804, 2010.

\bibitem{benitez2010}
P.~Ben{\'i}tez, J.~C. Mi{\~n}ano, and J.~C. Gonz{\'a}lez.
\newblock {\em Optics Express}, 18:7650, 2010.

\bibitem{minano2010}
J.~C. Mi{\~n}ano, P.~Ben{\'i}tez, and J.~C. Gonz{\'a}lez.
\newblock {\em New J. Phys.}, 12:123023, 2010.

\bibitem{smolyaninov2010}
I.~I. Smolyaninov and E.~E. Narimanov.
\newblock {\em Phys. Rev. Lett}, 105:067402, 2010.

\end{thebibliography}
\end{document}